\documentclass[sigconf,screen=true,bookmarks=false,9pt]{acmart}
\usepackage[linesnumbered,ruled,vlined]{algorithm2e}
\usepackage{graphicx} 
\usepackage{color} 
\usepackage{caption} 
\usepackage{tabularx}

\usepackage{fancyhdr}
\usepackage{amsmath,amsfonts}
\usepackage{algorithmic}
\usepackage{graphicx}
\usepackage{textcomp}
\usepackage{xcolor}
\usepackage{hyperref}
\usepackage{colortbl}
\usepackage{multirow}
\usepackage{footnote}
\usepackage{tablefootnote}
\usepackage{float}
\usepackage{threeparttable}
\usepackage{booktabs}
\usepackage{amsmath}
\hypersetup{
    colorlinks=true,
    linkcolor=blue,
    filecolor=magenta,      
    urlcolor=cyan,
    pdftitle={Overleaf Example},
    citecolor=blue,
}
\AtBeginDocument{%
  \providecommand\BibTeX{{%
    \normalfont B\kern-0.5em{\scshape i\kern-0.25em b}\kern-0.8em\TeX}}}

\def\BibTeX{{\rm B\kern-.05em{\sc i\kern-.025em b}\kern-.08em
    T\kern-.1667em\lower.7ex\hbox{E}\kern-.125emX}}
    
\definecolor{Gray}{gray}{0.85}

\newcommand{\method}{FastQuery}
\newcommand{\Dec}{\mathrm{Dec}}
\newcommand{\Enc}{\mathrm{Enc}}

\copyrightyear{2024} 
\acmYear{2024} 
\setcopyright{acmlicensed}\acmConference[DAC '24]{61st ACM/IEEE Design
Automation Conference}{June 23--27, 2024}{San Francisco, CA, USA}
\acmBooktitle{61st ACM/IEEE Design Automation Conference (DAC '24), June
23--27, 2024, San Francisco, CA, USA}
\acmDOI{10.1145/3649329.3657374}
\acmISBN{979-8-4007-0601-1/24/06}

\begin{document}

\title{FastQuery: Communication-efficient Embedding Table Query for Private LLM Inference}

\author{
    Chenqi Lin$^{1\dag}$,
    Tianshi Xu$^{1\dag}$,
    Zebin Yang$^{1\dag}$,
    Runsheng Wang$^{134}$, Ru Huang$^{134}$and Meng Li$^{213*}$
}
\affiliation{%
  \institution{$^1$School of Integrated Circuits \& $^2$Institute for Artificial Intelligence, Peking University, China}
  \country{}
}
\affiliation{%
  \institution{$^3$Beijing Advanced Innovation Center for Integrated Circuits, Beijing, China}
  \country{}
}
\affiliation{%
  \institution{$^4$Institute of Electronic Design Automation, Peking University, Wuxi, China}
  \country{}
}
\affiliation{%
  \institution{$^\dag$Equal contribution, $^*$Corresponding author: meng.li@pku.edu.cn}
  \country{}
}
\begin{abstract}

    With the fast evolution of large language models (LLMs), privacy concerns with user queries arise 
    as they may contain sensitive information. Private inference based on homomorphic encryption (HE)
    has been proposed to protect user query privacy. However, a private embedding table query has to
    be formulated as a HE-based matrix-vector multiplication problem and suffers from enormous
    computation and communication overhead. We observe the overhead mainly comes from the neglect of
    1) the one-hot nature of user queries and 2) the robustness of the embedding table to low bit-width quantization noise.
    Hence, in this paper, we propose a private embedding table query optimization framework, dubbed \method.
    \method~features a communication-aware embedding table quantization algorithm and a one-hot-aware
    dense packing algorithm to simultaneously reduce both the computation and communication costs.
    Compared to prior-art HE-based frameworks, e.g., Cheetah, Iron, and Bumblebee, FastQuery achieves 
    more than $4.3\times$, $2.7\times$, $1.3\times$ latency reduction, respectively 
    and more than $75.7\times$, $60.2\times$, $20.2\times$ communication reduction, respectively, on both LLAMA-7B and LLAMA-30B.
    
\end{abstract}

\maketitle
\pagestyle{fancy}
\section{Introduction}
\label{intro}

Large language models (LLMs), such as ChatGPT \cite{brown2020chatgpt} and LLAMA \cite{touvron2023llm_llama_model},
have demonstrated state-of-the-art performance in a wide range of tasks 
and get increasingly adopted in sensitive and private applications, such as 
personal assistants, medical diagnosis, etc.
However, to leverage LLMs on the cloud, users are required to disclose their
queries directly to the service provider, which may contain
sensitive information. Privacy has thus emerged as a major concern \cite{ren2023cham}.

\begin{figure}[!tb]
    \centering
    \includegraphics[width=0.9\linewidth]{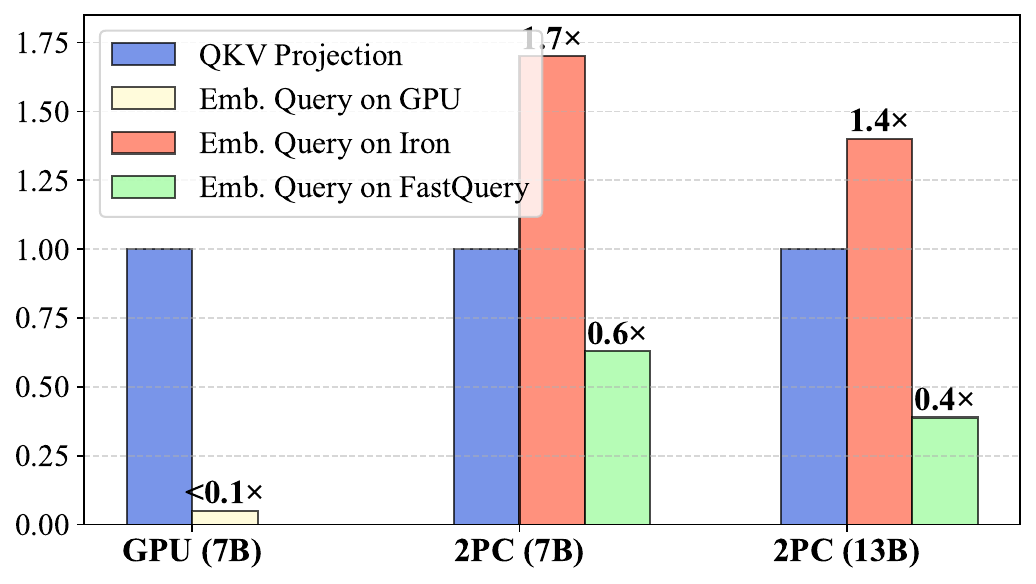}
    \caption{Compare the latency of QKV projection with embedding table query on GPU and prior-art 2PC framework Iron as well as our 2PC framework~\method~on LLAMA-7B and 13B. The data is normalized where the QKV projection is $1.0$. 
    }
    \label{fig:intro}
\end{figure}

To address these privacy concerns, secure two-party computation (2PC) based on homomorphic encryption (HE) has been proposed recently 
as a promising solution \cite{huang2022cheetah,hao2022iron}.
Besides enabling accurate LLM inference, it also protects the privacy of both user data and LLM model parameters with a formal
privacy guarantee \cite{lu2023bumblebee,hou2023ciphergpt,ren2023cham,hao2022iron}.

However, the privacy protection of HE-based 2PC is achieved at the cost of high communication overhead.
This is due to the transfer of a large number of 
high bit-width homomorphically encrypted ciphertexts
(e.g., 109-bit ciphertext for Cheetah \cite{huang2022cheetah})
between the server and the client. Previous works, such as Iron~\cite{hao2022iron}, CipherGPT~\cite{hou2023ciphergpt},
and Bumblebee~\cite{lu2023bumblebee}, etc, have focused on
optimizing the HE-based Transformer computation.
However, they overlook the private embedding 
table query, which we have found to be more time-consuming 
and communication-intensive 
compared to a transformer block. As depicted in Figure~\ref{fig:intro},
while the cost of the embedding table query is negligible in plaintext,
it incurs significantly higher latency in ciphertext compared to the QKV projection.
This is because to protect the user query privacy, embedding table query has to be realized with
a HE-based matrix-vector multiplication \cite{lu2023bumblebee}.
Considering the vocabulary size of LLM is very large, e.g., 32000 for LLAMA-7B, the communication overhead is massive and
thus the latency becomes substantial.

To enable efficient private embedding table query, we observe previous works ignore the key characteristics of embedding table.
On the one hand, the query is a one-hot vector since each time only one token from the vocabulary is picked. On the other hand,
the embedding table is robust against low bit-width quantization noise. For example, we empirically find directly quantizing the
embedding table to 6 bits introduces negligible perplexity degradation (e.g., 7.404 vs 7.340 for LLAMA-7B).
However, all existing works assume high-precision
plaintext for the embedding table \cite{huang2022cheetah,hao2022iron,lu2023bumblebee},
leading to a significant waste of communication. To this end, we propose \method, a
communication-efficient embedding table query protocol. 
By leveraging the one-hot nature of user queries and a low bit-width quantized embedding table, 
\method~significantly reduces the communication overhead.
The contributions of this work can be summarized as follows:
\begin{itemize}
    \item We identify that existing HE-based 2PC frameworks overlook the optimization opportunities for embedding table queries and propose \method,
        a communication-efficient protocol customized for private embedding table query.
    \item We introduce a communication-aware embedding table quantization algorithm to reduce the bit-widths of ciphertexts.
        Additionally, we develop a novel HE-based query protocol that utilizes the one-hot nature of user queries,
        further reducing communication overhead.
    \item \method~outperforms prior-art HE-based 2PC frameworks, including Cheetah, Iron, and Bumblebee.
        On LLAMA-7B and LLAMA-13B models, \method~achieves  more than $75.7\times$, $60.2\times$, $20.2\times$ communication reduction respectively   
        with negligible model performance degradation.
\end{itemize}

\section{Preliminary}
\label{sec:prelim}


\subsection{Threat Model}
\label{subsec:threat_model}

\method~focuses on efficient privacy-preserving DNN inference involving 
two parties, i.e., server and client.
The server holds the model with private weights
and the client holds private inputs. The model architecture,
including the number of layers as well as the types, dimensions,
and bit-widths for each layer, are public to both the server and the client
\cite{huang2022cheetah,hao2022iron,lu2023bumblebee}.
After executing the protocol, the client learns the inference results
without revealing any information to the server.
Consistent with previous works 
\cite{hao2022iron,lu2023bumblebee,huang2022cheetah}, 
we adopt an honest-but-curious
security model in which both parties follow the specifications of
the protocol but also try to learn more from the information than allowed.

\subsection{Flow of Secure Embedding Table Query}
\label{subsec:overview}

\begin{figure}[!tb]
    \centering
    \includegraphics[width=1.0\linewidth]{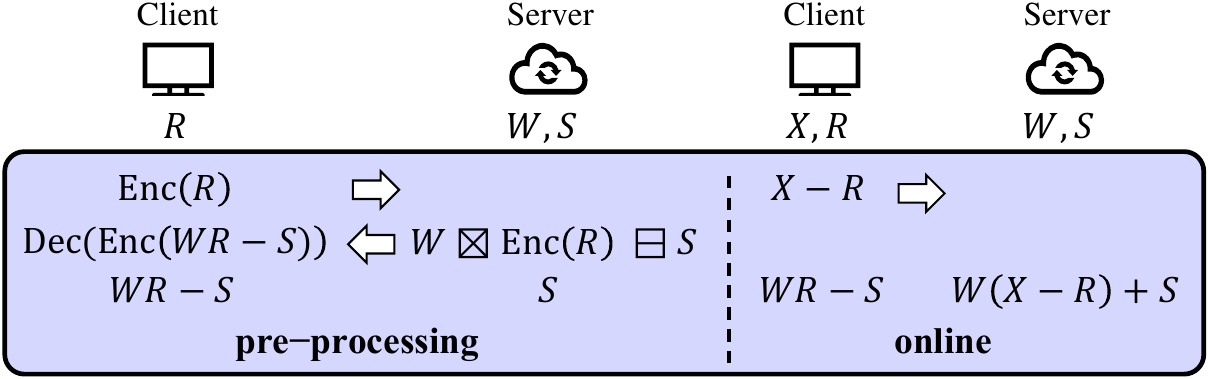}
    \caption{Flow of secure embedding table query. 
    }
    \label{fig:pre_HE}
\end{figure}


We summarize the notations used in the paper in Table~\ref{tab:notation}.
Figure~\ref{fig:pre_HE} shows the flow of secure embedding table query. Previous work \cite{Mishra_Delphi_2020} considers embedding table query as a matrix-vector multiplication, e.g.,  $WX$, which has two stages: the pre-processing stage that generates the input-independent helper data and the online stage to process the client's real query. During the pre-processing stage, the client encrypts a randomly sampled $R$ and sends it to the server. Since HE supports homomorphic addition and multiplication without the need for decryption, the server can compute $\Enc(WR)$ and blind it by subtracting a randomly sampled $S$. Finally, the server sends the result to the client for decryption. During the online stage, the client blinds a one-hot query $X$ by subtracting $R$, i.e., $X-R$, and sends it to the server. Then, the server just needs to compute $W(X-R)+s$ and each party gets a secret share of the results $WX$, i.e., $WR - S$ by the client and $W(X-R) + S$ by the server.

\begin{table}[!tb]
    \centering
    \caption{Notations used in the paper.}\label{tab:notation}
    \resizebox{1.0\linewidth}{!}{
        \begin{tabular}{c|c}
        \toprule
        Notations & Meanings \\
        \midrule
        $\lceil \cdot \rceil$,$\lfloor\cdot \rfloor$,$\lfloor\cdot \rceil$ & Ceiling, flooring, and rounding operations \\
        \hline
        $\Enc(\cdot), \Dec(\cdot)$ & Homomorphic encryption and decryption\\
        \hline
        $\boxplus,\boxminus,\boxtimes $  & Homomorphic addition, subtraction and multiplication  \\
        \hline
        \multirow{1}{*}{$N, p,q$} & Polynomial degree, bit-width of plaintext and ciphertext\\
        \hline
        $m,n$ & Vocabulary size and embedding dimension \\ 
        \bottomrule
        \end{tabular}
    }
\end{table}

\subsection{Communication Complexity of Embedding Table Query}
\label{subsec:comm_cost}

The main communication cost of embedding table query comes from the pre-processing stage while 
the online stage is usually lightweight since it only transfers $X-R$ in plaintext. 
The communication cost primarily comes from two parts in the pre-processing stage.
As shown in Figure~\ref{fig:pre_HE}, the first part is the transfer of input ciphertext $\Enc(R)$ from the
client to the server, which equals the product of the number of polynomials
and the communication of each polynomial. The number of polynomials is $\lceil \frac{m}{N} \rceil$,
and the communication of each polynomial scales with both the polynomial degree and
the bit-width of each coefficient. Hence, the input communication complexity is $O(\lceil \frac{m}{N} \rceil \times Nq)$.
The second part is the transfer of output ciphertext from the server to
the client. Following the optimization of Cheetah, 
the output transfer communication is $O(n/ \lceil \frac{N}{m}\rceil \times Nq + nq)$.
Since the size of embedding table $m$ is quite large, 
e.g. 32000 for LLAMA-7B, $\lceil \frac{N}{m}\rceil =1$ and thus the output communication can be simplified as $ O(n\times Nq+nq)$.


\noindent \textbf{HE parameters} In HE, the bit-width of plaintext $p$ is decided by the accumulation bit-width of a matrix-vector multiplication to avoid overflow. Given $p$,
$q-p$ determines the noise budget of the HE-based computation, which is related to the multiplication depth.
Given $q$, $N$ can be determined according to the required security level.

\noindent \textbf{PackLWEs optimization} To improve the packing density of output polynomials, \cite{lu2023bumblebee} proposes PackLWEs optimization which extracts the useful elements in each polynomial and packs them into a new polynomial. With the PackLWEs optimization, the number of output ciphertexts is reduced from $n$ to $\lceil \frac{n}{N}\rceil$. 

\subsection{Coefficient Packing of HE}
\label{subsec:coe_pack}

\begin{figure}[!tb]
    \centering
    \includegraphics[width=1.0\linewidth]{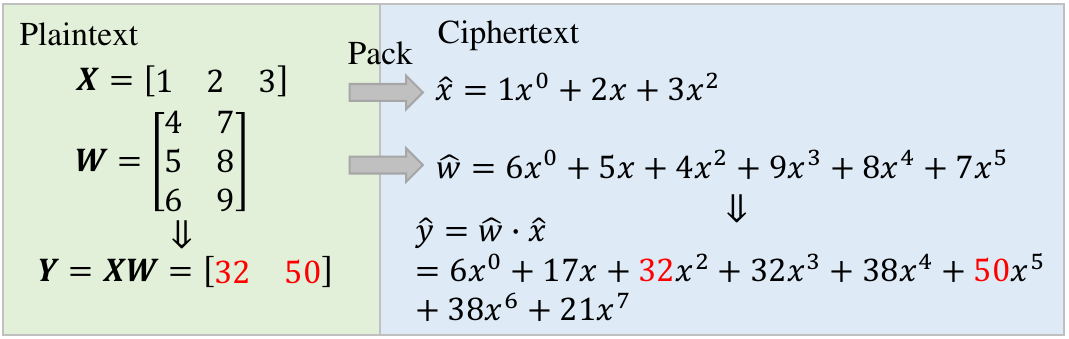}
    \caption{A matrix-vector multiplication example of coefficient packing.
    }
    \label{fig:coe_pack}
\end{figure}
HE encrypts plaintexts into ciphertext polynomials, namely, 1-dimensional vectors. However, DNNs operate on high-dimensional tensors. To bridge this gap, packing is necessary to encode tensors into polynomials. Cheetah~\cite{huang2022cheetah} recognizes that polynomial multiplications inherently perform matrix-vector multiplication and introduces a HE coefficient packing method. In this approach, tensor elements are encoded into coefficients, and the correct result can be extracted from the outcome of polynomial multiplication. Figure~\ref{fig:coe_pack} provides a toy example of this process.
Iron further optimizes the algorithm of Cheetah for matrix-matrix multiplication.
CHAM \cite{ren2023cham} adopts a different weight matrix partitioning method to maximize the number of output ciphertexts,
which enables fully utilizing PackLWEs to reduce communication overhead.

\subsection{Transformer Model Quantization}
\label{subsec:quantization}
Quantization is a frequently used method in model compression. There are two approaches for language model quantization, quantization-aware training (QAT) \cite{liu2023llm} and post-training quantization (PTQ) \cite{kim2023squeezellm}. For LLM quantization, PTQ is adopted by most of the works, due to the lack of access to training data and the high retraining cost of QAT \cite{kim2023squeezellm}. However, there is little work focusing on LLM embedding quantization. This is because most of the parameters in LLMs are distributed in the encoder layers rather than the embedding table. In this paper, however, we propose a communication-aware post-training embedding table quantization method to reduce embedding table query latency.


\section{Motivation}
\label{sec:motivation}


In this section, we analyze the origin of the communication cost associated with embedding table queries to motivate our work.
As discussed in Section~\ref{subsec:comm_cost}, the communication cost for input and output ciphertexts is given by $O(\lceil \frac{m}{N} \rceil \times Nq)$ and $O(n/ \lceil \frac{N}{m}\rceil \times Nq +nq)$, respectively. We have made several observations to reduce the communication further.


\textbf{Observation 1: Both HE parameters $p$ and $q$ can be reduced with low bit-width quantized embedding query, bringing in improved communication efficiency.} As discussed in Section~\ref{subsec:comm_cost}, the bit-width of plaintext $p$ and ciphertext $q$ is determined by the accumulation bit-width of a matrix-vector multiplication. Low bit-width quantization directly reduces the required accumulation bit precision, leading to efficient communication. Furthermore, the robustness of DNNs to quantization error suggests a promising direction for efficiency improvement.

\textbf{Observation 2: Although the embedding table can be quantized to low precision, e.g., sub-4-bit quantization, further reduction of $p$ and $q$ is not feasible.} Previous research has demonstrated the feasibility of quantizing LLM weights to low bit-widths, such as 3 bits~\cite{kim2023squeezellm,frantar2022gptq}. However, we have empirically found that reducing $p$ below 13 bits compromises the correctness of the decryption. Hence, the communication reduction saturates once $p$ is reduced below 13 bits~\cite{sealcrypto,xu2024hequant}.
Therefore, to fully leverage LLMs' robustness to low bit-width quantization, how to reduce communication for sub-13-bit plaintext becomes important.

\textbf{Observation 3: Existing approaches move the HE computation to the pre-processing stage and thus, cannot leverage the one-hot nature of user queries.}
Previous works~\cite{hao2022iron,hou2023ciphergpt,huang2022cheetah} move the HE computation to the pre-processing stage, where we compute $WR$ instead of $WX$. However, since the query $X$ is a one-hot vector with only one non-zero element, there are possibilities to further reduce the communication cost.
Furthermore, recent studies suggest that for DLaaS, pre-processing communication cannot be ignored and often incurs online latency due to insufficient server downtime to hide the communication costs \cite{garimella2023characterizing}. Therefore, computing $WX$ 
directly in the online stage is a promising direction for further reducing the total communication cost.



\section{FASTQUERY}
\label{sec:method}

\subsection{Overview}
\label{subsec:overview}

\begin{figure}[!tb]
    \centering
    \includegraphics[width=1.0\linewidth]{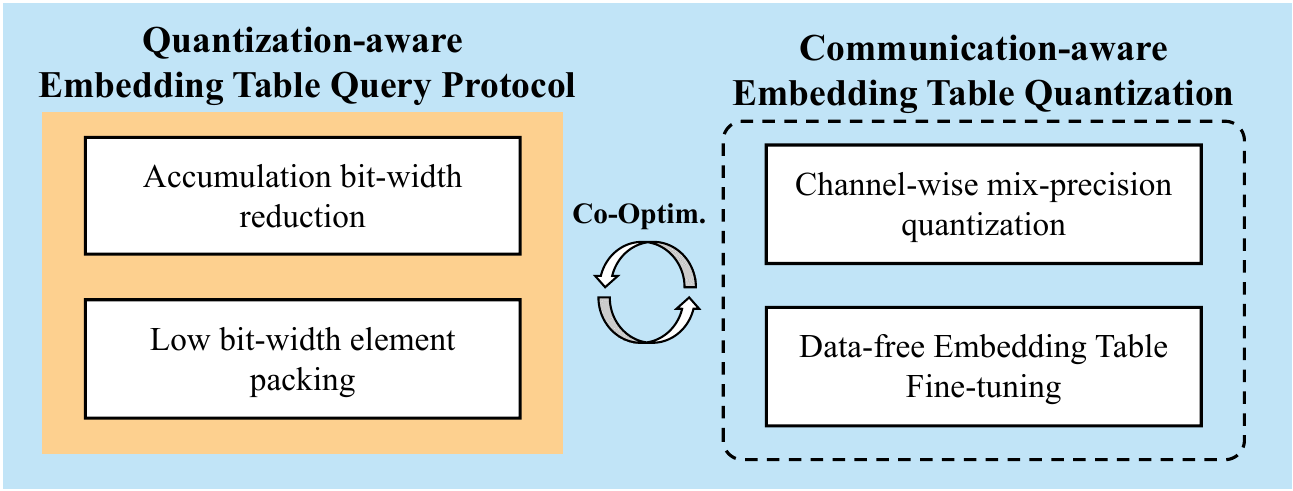}
    \caption{Overview of FastQuery Framework. 
    }
    \label{fig:framework}
\end{figure}

Based on the observations above, we propose~\method, as shown in Figure~\ref{fig:framework}. 
\method~features both network and protocol optimizations.
For protocol optimization, we leverage the one-hot nature of user queries and the low bit-width embedding table
to reduce the accumulation bit-width, and thus, the plaintext and ciphertext bit-widths.
To further reduce the communication for sub-13-bit quantization, we propose a novel
element packing algorithm to squeeze the embedding table dimension.
For network optimization, we propose a channel-wise mix-precision quantization method
to reduce the bit-width of the embedding table, which also considers the proposed element packing protocol.
After quantization, we adopt a data-free fine-tuning to further recover model performance.
With the network and protocol co-optimization, \method~demonstrates over 10$\times$ communication reduction
compared to baseline protocols.

\subsection{Quantization-aware Embedding Table Query Protocol}


\begin{figure}[!tb]
    \centering
    \includegraphics[width=1.0\linewidth]{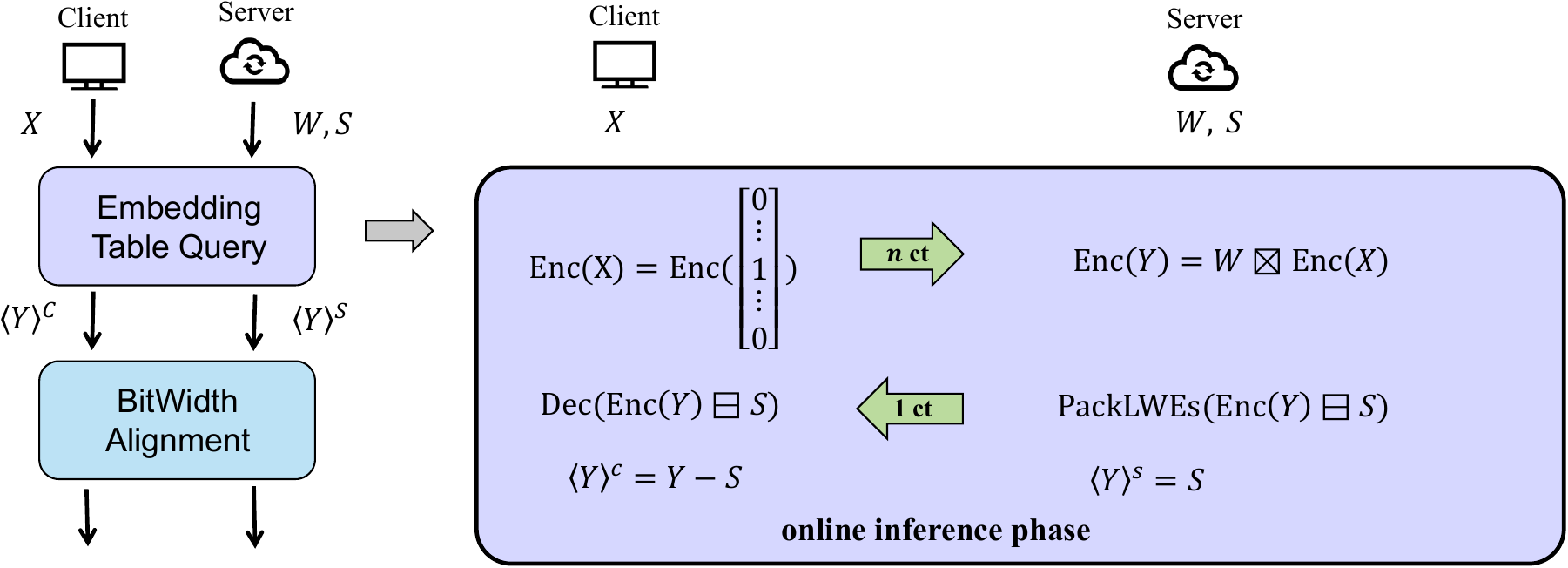}
    \caption{The proposed private embedding table query protocol.}
    \label{fig:query_protocol}
\end{figure}

As mentioned in Section ~\ref{sec:motivation}, offline pre-processing communication not only can potentially waste power for both the server and the client but also can slow down the online stage. 
Therefore, we choose to perform the matrix-vector multiplication solely in the online phase.
Moreover, we propose several optimizations to fully utilize the characteristics of one-hot encoding to improve communication efficiency. In Figure \ref{fig:query_protocol}, we illustrate the proposed protocol for private embedding table query. 



\textbf{Accumulation bit-width reduction }
The core computation in the embedding table query is the matrix-vector multiplication $WX - S$.
When querying the embedding table, we need to compute $WX - S$ homomorphically. 
Let $b_x$ and $b_w$ denote the bit-widths of $X$ and $W$, respectively.
Then, a naive estimation of the bit-width for output vector $u$ is $b_u = b_x + b_w + b_{acc} + 1$,
where $b_{acc} = \log_2 m$ is the bit-width required for the accumulation of the matrix-vector multiplication
(recall $m$ is the vocabulary size)
and $+1$ is to ensure the randomly sampled $S$ can be corrected subtracted~\cite{huang2022cheetah}.
To guarantee HE computation correctness, we need to have $p \leq b_u$.
However, we find this is not necessary. Because the query is a one-hot vector,
it is sufficient for $p = b_u = b_w + 1$ since there is only one non-zero value, which equals 1. 
Therefore, we can significantly reduce the plaintext modulus and the communication cost.


\begin{figure}[!tb]
    \centering
    \includegraphics[width=1.0\linewidth]{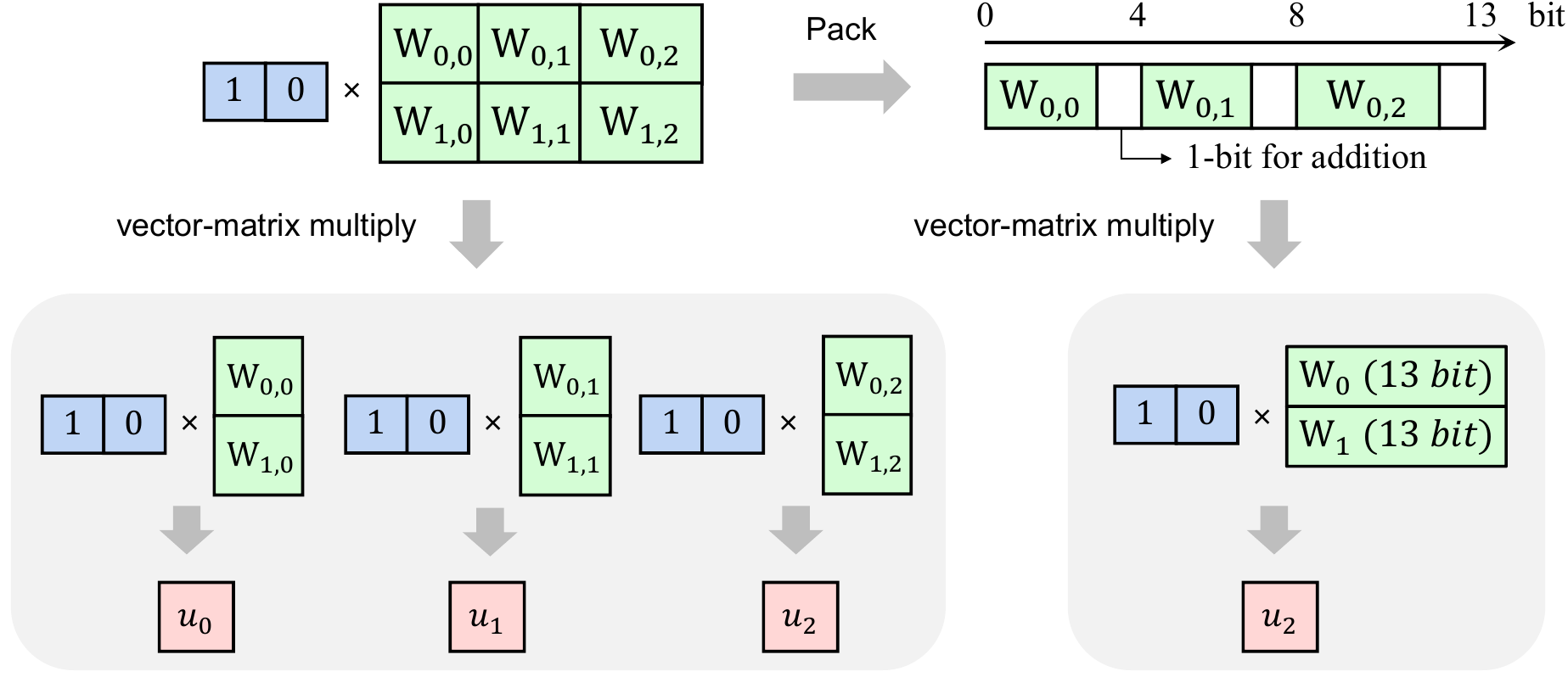}
    \caption{Toy example of how we pack a matrix with three columns into one column. The bit-widths for each element in the matrix's channels are 3, 3, and 4 respectively and every 13-bit plaintext polynomial coefficient is fully packed. By employing the method we propose, we can merge the three channels into a single channel.
    }
    \label{fig:pack_method}
\end{figure}

\textbf{Low bit-width aware element packing }
While the one-hot nature of user queries enables us to reduce the accumulation bit-width to $b_w + 1$,
the plaintext bit-width $p$ cannot be reduced to below 13 bits to ensure correct decryption as introduced in Section~\ref{sec:prelim}.
Hence, it is important to propose further optimizations for communication reduction.
Specifically, we can pack multiple low bit-width elements into an element with a higher precision,
making full use of the high-precision plaintext coefficients. In~\method, we fix the plaintext bit-width to 13. 
Depending on the embedding table precision, we always try to pack as many elements as possible.
With the proposed packing algorithm, each dot product between the input vector and a column is equivalent to
performing a dot product with multiple embedding table columns, 
which reduces both the computation and number of output polynomials for better communication
efficiency. In Figure~\ref{fig:pack_method}, we illustrate the packing process and provide an example to demonstrate the benefits
of matrix-vector multiplication.



\textbf{Bit-width alignment }
After computing the HE-based matrix-vector multiplication, we need to align the bit-width of each output element before the computation of the next layer.
We extend the bit-width of all elements to the required bit by leveraging the bit-width extension protocol proposed by SiRNN \cite{rathee2021sirnn}.
The bit-width alignment step enables us to support a mixed-precision quantized embedding table, which 
boosts the performance of low bit-width quantized networks with negligible communication cost.




\subsection{Communication-aware Embedding Table Quantization}

\begin{table}[!tb]
    \centering
    \caption{Comparison on different granularities of INT3 embedding table quantization, evaluated on WikiText-103. Per-token quantization is forbidden as it can leak the query information.
    }
    \label{tab:exp:quantization granularity}
    \scalebox{0.8}{
    \vspace{3pt}
    \begin{tabular}{l|c|c|c}
    \toprule
    \textbf{Quantization Granularities}  &\textbf{LLaMA-7B} & \textbf{LLaMA-13B}&\textbf{LLaMA-30B} \\ 
    \midrule
    Per-tensor             &15236.57     & 11210.18    &1772.56    \\
    \midrule
    Per-token (\textbf{data leakage})  &10.949      & 9.576    &6.000    \\
    \midrule
    Per-channel            &10.950      & 8.268    &5.659    \\
    \bottomrule
    \end{tabular}}
\end{table}


We now aim to reduce the bit-width of the embedding table while preserving the performance of LLMs.
While per-tensor quantization is easy to implement, we find it can hurt model performance,
as shown in Table \ref{tab:exp:quantization granularity}.
Since our proposed protocol can support mixed bit-widths, we study more fine-grained quantization strategies,
including per-token quantization and per-channel quantization.
As shown in Table~\ref{tab:exp:quantization granularity}, by increasing the quantization flexibility and the quantization parameters,
both per-token and per-channel quantization achieve significantly better performance.
However, per-token quantization can leak the query information.
Since the quantization parameters, including the scale and bit-width, are publicly known, an adversary can infer the query
based on which bit-width or scale is selected after the private table query.
Hence, we propose a per-channel quantization strategy that consists of the following three steps.

\textbf{Number of channels and bit-width combination in each coefficient}
As mentioned before, the bit-width of each plaintext coefficient is fixed to 13. In addition, to generate secret shares, the output needs to be subtracted by a random number $S$. So we need to reserve an additional 1 bit for each channel in each coefficient. Based on this, now we determine how many channels we should pack in one coefficient. If we set this number too low, e.g., two channels, the total number of coefficients will be high, leading to a higher embedding table query latency. If we set the number too high, e.g., four channels, more than half of the channels will be quantized to 2-bit or lower. We find that this will badly hurt model performance, shown in Table \ref{tab:exp:wiki-different salient weight strategy and } and Table \ref{tab:exp:different salient weight strategy and }.
So for query efficiency and model performance, we set the number of channels in each coefficient to 3. The bit-widths of the three channels are set to 4-bit, 3-bit, and 3-bit (represented by 4, 3, 3) to prevent 2-bit quantization, which may hurt model performance.

\begin{figure}[!tb]
    \centering
    \includegraphics[width=0.9\linewidth]{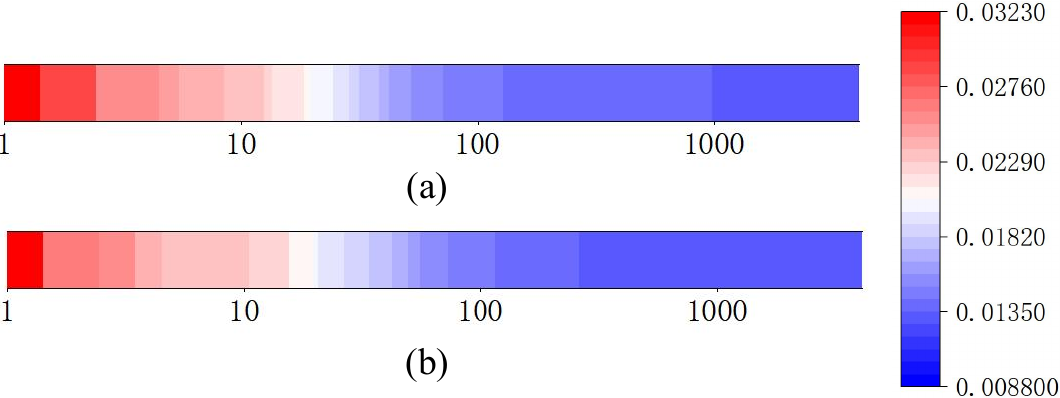}
    \caption{Mean absolute values of different embedding channels in (a) LLaMA-7B and (b) LLaMA-13B. }
    \label{fig:channel_difference}
\end{figure}

\textbf{Precision of each packing channel}
Now we need to determine which channel we should quantize to 4-bit in each coefficient. Considering outliers (weights with high absolute values) cause huge approximation errors in quantization, we observe the weight distribution in the embedding table and find the mean absolute value of each channel varies greatly, as shown in Figure \ref{fig:channel_difference}. Based on this, we quantize the salient channels whose weights have higher mean absolute values to 4-bit. There are other strategies such as considering gradient value \cite{lin2023awq} or Hessian value \cite{frantar2022gptq,kim2023squeezellm}. Our method achieves the best performance in this bit-width combination and is easy to implement, as shown in Table \ref{tab:exp:wiki-different salient weight strategy and } and Table \ref{tab:exp:different salient weight strategy and }.

In practice, we just quantize $1/3$ of the embedding channels that have the highest mean absolute value into 4-bit and the other $2/3$ into 3-bit. Then we change the order of embedding channels and the order of QKV projection weight rows in the first MHA block together, keeping the multiplication result unchanged. At last, the bit-widths of three adjacent channels will be 4-bit, 3-bit, and 3-bit, which can be packed in a coefficient.
This adjustment can unify the bit-width combination order in each coefficient and will not introduce any calculation error.

\begin{figure}[!tb]
    \centering
    \includegraphics[width=0.8\linewidth]{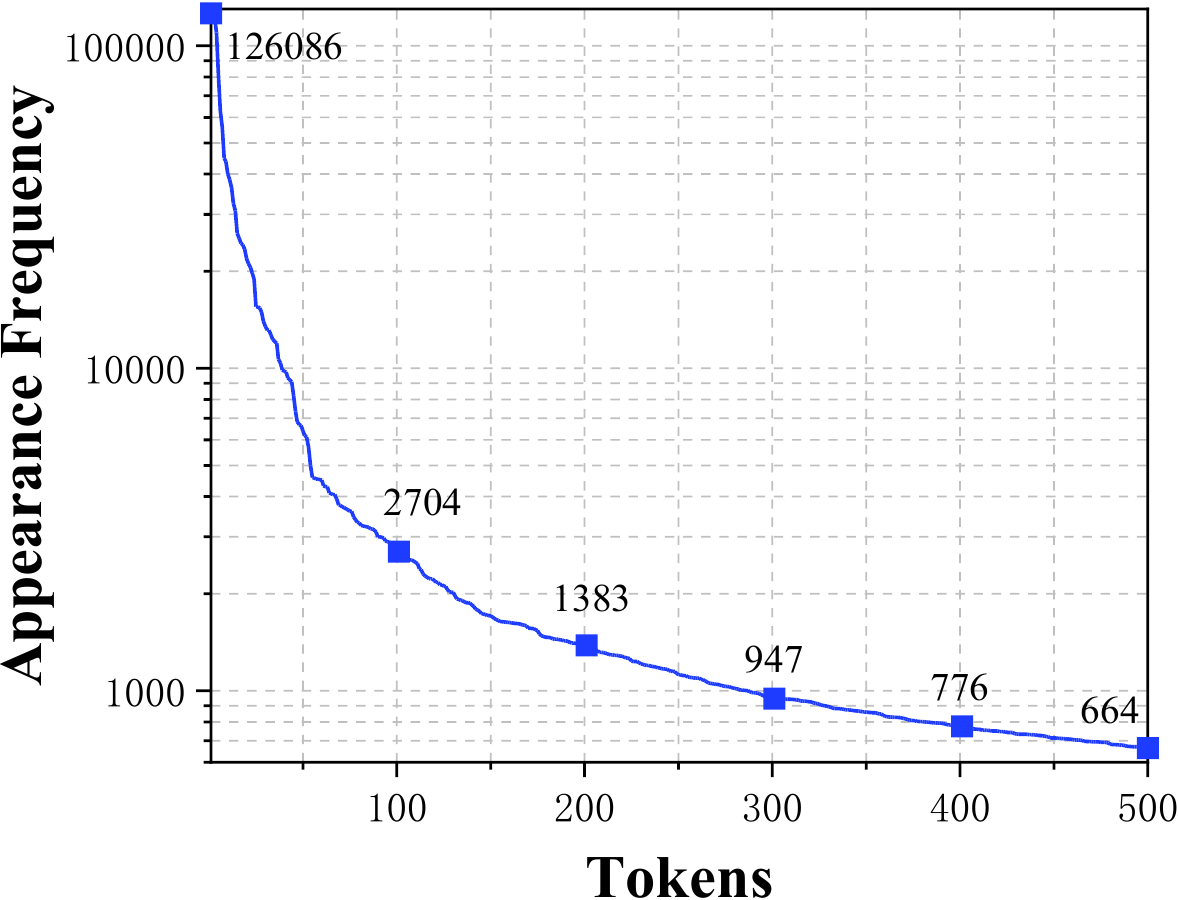}
    \caption{Appearance frequencies of the most frequent 500 tokens in the WikiText-103 dataset.}
    \label{fig:frequency}
\end{figure}

\textbf{Data-free Embedding Table Fine-tuning} 
To further improve the model performance, we need to conduct fine-tuning on the quantized embedding table. A simple idea is to train the model on the downstream task, while this will lead to huge training costs and overfitting problems \cite{kim2023squeezellm}. So we need a data-free post-training algorithm to conduct fine-tuning.
At the same time, we find that different tokens in vocabulary usually have different importance because their appearance frequencies vary greatly, as shown in Figure \ref{fig:frequency}.
Based on these, we design a data-free embedding table fine-tuning algorithm to further reduce the impact of quantization error. In this period, we want to minimize 
\begin{align*}
    L=\Vert I_{token} (Q(W'_{Emb})W_{QKV} - W_{Emb}W_{QKV}) \Vert_2,
\end{align*}
where $W_{QKV}$ is the concatenation of Q, K, and V projection weight in the first MHA block. $W_{Emb}$ and $Q(W'_{Emb})$ is 
the embedding table before and after quantization. $Q()$ is a function representing our per-channel mix-precision quantization method. 
$I_{token}$ is a diagonal matrix and the elements on the diagonal represent the appearance frequency of each token. This can better protect embedding vectors for important tokens that have a higher frequency. In practice, We set a threshold value to lower the elements in $I_{token}$ which is bigger than this value to this value. This can prevent the fine-tuning algorithm from paying too much attention to only a few tokens.
We freeze $I_{token}$, $W_{Emb}$, and $W_{QKV}$, only changing $Q(W'_{Emb})$ with gradient descent and straight-through estimator (STE)
. 
This further improves model performance without introducing much training time.
It is worth noting that this fine-tuning algorithm is completely data-free, preventing the problem of over-fitting on downstream tasks faced by existing LLM quantization methods.

\section{EXPERIMENT RESULTS}
\label{sec:exp}


\subsection{Experimental Setup}
FastQuery is built on top of SEAL library \cite{sealcrypto} and OpenCheetah \cite{huang2022cheetah} in C++.
We use EZPC \cite{chandran2017ezpc} library to evaluate CrypTFlow2 \cite{rathee2020cryptflow2} and SIRNN \cite{rathee2021sirnn} and use SPU library \cite{spu} to simulate Iron \cite{hao2022iron} and Bumblebee \cite{lu2023bumblebee}.
All experiments were conducted on a server with a 2.4GHz Intel Xeon CPU and 125GB RAM.
Following the Cheetah experiment protocol, we set $N$ to 4096 for most of our experiments.
We quantize and evaluate LLaMA-2 family \cite{touvron2023llama} on WikiText \cite{merity2016pointer} and
C4 \cite{raffel2020exploring}, using a symmetric uniform quantizer. We round each 
scale factor to the power of 2 to be friendly to 2PC protocols \cite{rathee2021sirnn}.

\subsection{Benchmark for Embedding Table Quantization}

\textbf{Perplexity and Query Latency Comparison}
As we are the first to focus on LLM embedding table quantization, we compare FastQuery with Round to Nearest per-channel quantization (RTN) \cite{yao2022zeroquant}. As shown in Table \ref{tab:exp:accuracy_comp_other_KD}, RTN INT4 quantization shows similar perplexity with the full precision model, while RTN INT3 can cause an obvious model performance drop.
Compared with RTN INT4, FastQuery can pack more channels in one coefficient, which leads to a smaller number of coefficients with only a small performance drop. This can bring $1.39 \sim 1.52 \times$ query latency reduction compared to RTN INT4 and $2.95 \sim 3.40 \times$ compared to full precision models. Compared with RTN INT3, with the same query latency, FastQuery achieves much lower perplexity on each model size. After data-free fine-tuning, the performance of the quantized model can be further improved. On WikiText-103, our quantized LLaMA-13B and LLaMA-30B model even achieves lower perplexity than RTN INT4.

\begin{table}[!tb]
    \centering
    \caption{Comparison on WikiText-103 and C4 perplexity (↓) and embedding table query latency.}
    \label{tab:exp:accuracy_comp_other_KD}
    \scalebox{0.7}{
    \vspace{3pt}
    \begin{tabular}{lcc|c|c|c|c|c|c}
    \toprule
    \multirow{2}*{Method}&\# channels &\multirow{2}*{Metrics}  &\multicolumn{2}{c|}{LLaMA-7B} & \multicolumn{2}{c|}{LLaMA-13B} & \multicolumn{2}{c}{LLaMA-30B} \\ 
                           &   per-coefficient        &              & Wiki      &C4        & Wiki       &C4        & Wiki       &C4 \\
    \midrule
    \multirow{2}{*}{FP16}  &\multirow{2}{*}{/}  & PPL (↓)     &7.340      & 8.220    &7.313       &7.281     &    5.576   &6.704    \\
                          &     & Latency (s)                 &   11.39  &   11.39       &    14.27        &    14.27      &     17.45       &  17.45 \\
    \midrule
    RTN&\multirow{2}{*}{2}&PPL (↓)                            & 7.403     & 8.236    &7.326       &7.284     &   5.615    &6.706   \\
     (INT4)                        &  & Latency (s)               &   5.11        &     5.11     &       6.39     &     6.39
     &    8.23        &  8.23  \\
    \midrule
   RTN&\multirow{2}{*}{3}&PPL (↓)                             & 10.950    & 17.738   &8.268       &12.390    &  5.659     &6.881    \\
     (INT3)                       &   & Latency (s)               &  3.35         &   3.35       &      4.33      &    4.33      &     5.91       & 5.91  \\
    \midrule
    \multirow{2}{*}{FastQuery}&\multirow{2}{*}{3} &PPL (↓)    &7.679      &  8.409   & 7.327      & 7.312    & 5.618      & 6.713   \\
                            &   & Latency (s)                      &  3.35         &   3.35       &      4.33      &    4.33      &     5.91       & 5.91  \\
    \midrule
    FastQuery                &\multirow{2}{*}{3}  & PPL (↓)   &7.430      &  8.285   & 7.317      &7.303     & 5.600      & 6.713   \\
    +Fine-tuning            &    & Latency (s)                      &  3.35         &   3.35       &      4.33      &    4.33      &     5.91       & 5.91  \\
    \bottomrule
    \end{tabular}}
\end{table}

\textbf{Different bit-width combinations and salient channel judgment strategies}
In our method, we use the mean absolute value to select the salient channels. In each 13-bit coefficient, we use the bit-width combination of 4-bit, 3-bit, and 3-bit (represented by 4, 3, 3). We also evaluate other bit-width combinations and salient channel judgment strategies, such as mean gradient value, mean Hessian value, etc. We find our strategy achieves the best performance on various datasets, as shown in Table \ref{tab:exp:wiki-different salient weight strategy and } and Table \ref{tab:exp:different salient weight strategy and }. We also find that 2-bit quantization should be avoided since even $1/3$ of the channels quantized to 2-bit can cause notable model performance reduction.

\begin{table}[!tb]
    \centering
    \caption{Comparison of different bit-width combinations and salient channel judgment, evaluated on WikiText-103 perplexity (↓) of LLaMA-7B.}
    \label{tab:exp:wiki-different salient weight strategy and }
    \scalebox{1.0}{
    \vspace{3pt}
    \begin{tabular}{c|c|c|c}
    \toprule
    \textbf{Criterion}       &\textbf{ 6, 2, 2}   &\textbf{ 5, 3, 2}    & \textbf{ 4, 3, 3}   \\
    \midrule
    Random                   & 17015.25 & 132.992 & 12.554  \\
    \midrule
    Gradient                 &   2249.00 &  36.349  & 8.243   \\
    \midrule
    Hessian value            &2670.43   &  55.715 & 8.478   \\
    \midrule
    Gradient*Absolute Value  &   1317.64 & 24.819   & 8.411  \\
    \midrule
    Absolute Value           & 750.80   &  15.313  & \textbf{7.679}  \\
    \bottomrule
    \end{tabular}}
\end{table}

\begin{table}[!tb]
    \centering
    \caption{Comparison of different bit-width combinations and salient channel judgment, evaluated on C4 perplexity (↓) of LLaMA-7B.}
    \label{tab:exp:different salient weight strategy and }
    \scalebox{1.0}{
    \vspace{3pt}
    \begin{tabular}{c|c|c|c}
    \toprule
    \textbf{Criterion}       &\textbf{ 6, 2, 2}   &\textbf{ 5, 3, 2}    & \textbf{ 4, 3, 3}   \\ 
    \midrule
    Random                   & 21308.77   &90.453     &  11.214   \\
    \midrule
    Gradient                 & 1008.85  &  11.450   &  8.439   \\
    \midrule
    Hessian value            &  1197.39  &12.468     &  8.441   \\
    \midrule
    Gradient*Absolute Value  &  591.41  &   10.060  &  8.432   \\
    \midrule
    Absolute Value           & 577.16   & 11.785   &  \textbf{8.409}   \\
    \bottomrule
    \end{tabular}}
    \vspace{-8pt}
\end{table}

\subsection{Benchmark for Embedding Table Query Protocol}
\begin{figure}[!tb]
  \centering
  \includegraphics[width=1.0\linewidth]{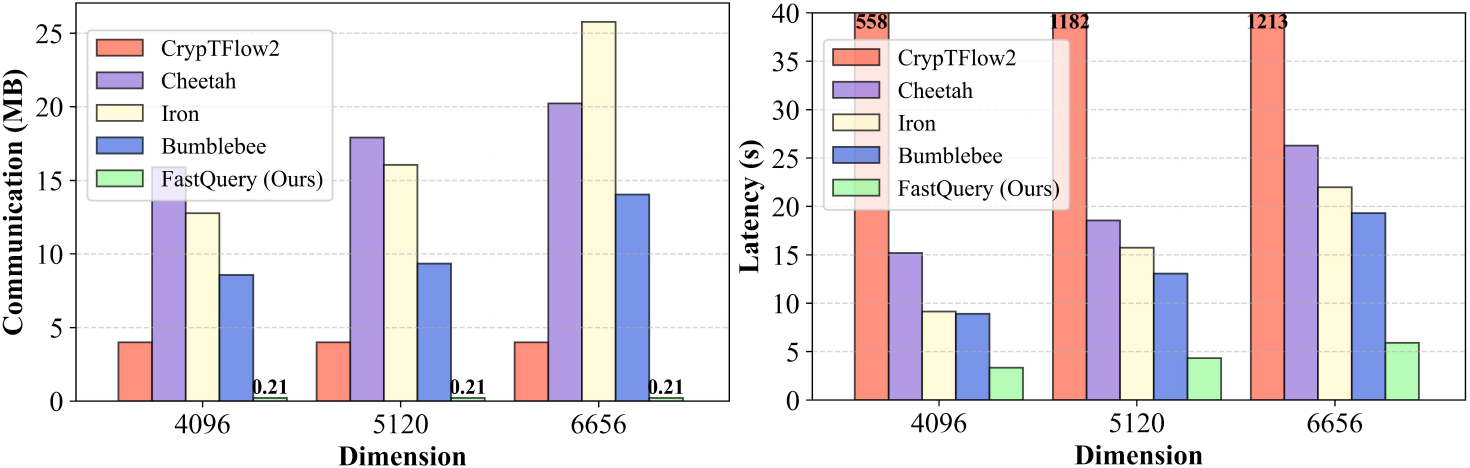}
  \caption{Communication (MB) and Latency (s) comparison under different embedding table dimensions.  
  }
  \label{fig:exp_micro}
  \vspace{-5pt}
\end{figure}

\textbf{Different dimensions evaluation}
We compare the communication of FastQuery with other methods across different embedding table dimensions. The experimental settings use embedding table dimensions from LLaMA-7B, LLaMA-13B, and LLaMA-30B. As shown in Figure~\ref{fig:exp_micro}, compared to Cheetah, FastQuery achieves a reduction in communication overhead of $75.2\sim 95.1\times$, as well as a reduction in latency of $4.3\sim 4.5\times$. FastQuery achieves a latency reduction of $2.7\sim 3.7\times$ compared to Iron. Compared to Bumblebee, FastQuery also achieves a latency reduction of $1.3\sim 1.5\times$ and a communication reduction of $20.4\sim 33.4\times$. Additionally, it can be observed that CryptFlow2 incurs significant latency overhead but relatively lower communication overhead. This is primarily due to its adoption of SIMD encoding, which reduces communication overhead through time-consuming rotations. In comparison, our method achieves a latency reduction of  $164.2\sim 205.2\times$ compared to CryptFlow2 and exhibits even lower communication overhead.

\begin{figure}[!tb]
  \centering
  \includegraphics[width=0.8\linewidth]{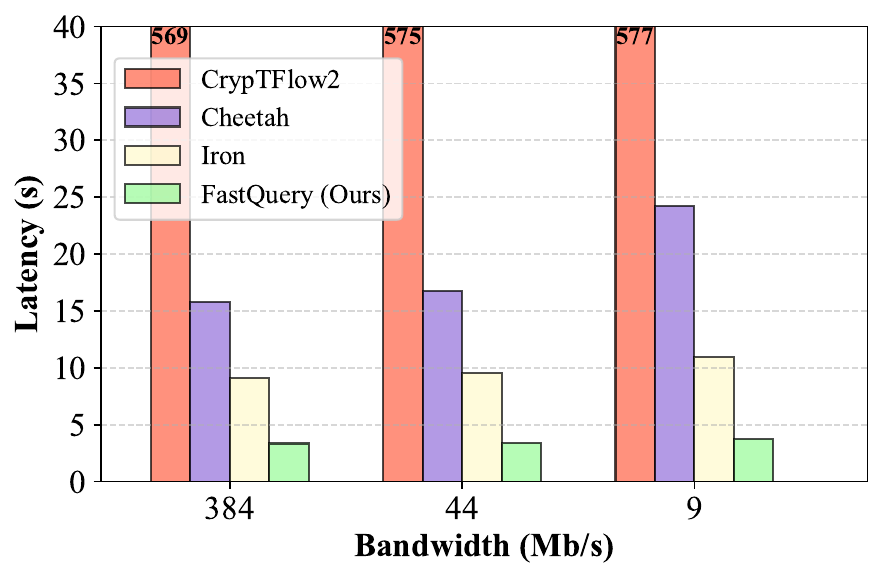}
  \caption{Lantency (s) comparison under different bandwidths. 
  }
  \label{fig:exp_bandwidth}
  \vspace{-5pt}
\end{figure}

\textbf{Different bandwidth evaluation}
We also compare the latency of FastQuery against CryptFlow2, Cheetah, and Iron under different bandwidths: 384Mb/s, 44Mb/s, and 9Mb/s, which are selected from the experiments of Cheetah\cite{huang2022cheetah} and Falcon\cite{xu2023falcon} respectively. As shown in Figure~\ref{fig:exp_bandwidth}, the latency changes of FastQuery and CryptFlow2 are not significant, while those of Cheetah and Iron are more evident. The experimental results are consistent with the fact that smaller communication overhead leads to less sensitivity to bandwidth. We also find that our method reduces the latency by $4.7\times$ to $6.5\times$ compared to Cheetah.

\subsection{Ablation study}
\begin{table}[!tb]
  \centering
    \caption{ablation study of our proposed method in FastQuery with a size of (32000, 4096).}
    \label{tab:ablation_study_of_protocol}
    \scalebox{0.8}{
    \begin{tabular}{cc|c|c} 
      \toprule
      \multicolumn{2}{c|}{\textbf{METHOD}} & \textbf{Latency (s)} & \textbf{Communication (MB)}\\
      \midrule
      \multicolumn{2}{c|}{Baseline Cheetah} & 15.19 & 15.933\\
      \midrule
      \multicolumn{2}{c|}{+Quant} & 15.19 & 15.933\\
      \midrule
      \multicolumn{2}{c|}{+Accumulation bit-width reduction} & 10.908 & 8.492\\
      \midrule
      \multicolumn{2}{c|}{+Low bit-width aware element packing}  & 3.492 & 4.911\\
      \midrule
      \multicolumn{2}{c|}{+PackLWEs} & 3.353 & 0.212\\
      \bottomrule
    \end{tabular}}
    \vspace{-5pt}
\end{table}
To better understand the improvements of different methods on communication efficiency and latency, we incrementally integrate our proposed optimization techniques into the baseline Cheetah and we present the results in Table \ref{tab:ablation_study_of_protocol}. According to the results, we can learn that 1) directly applying quantization does not provide significant benefits for accelerating embedding table query; 2) after quantization, reducing the bit-width of plaintext modulus and ciphertext modulus and packing low bit-width elements has a significant impact on reducing communication overhead and latency; 3) although the PackLWEs method incurs heavy computation costs, it reduces the communication significantly.
These experimental results demonstrate that the proposed methods are highly effective in reducing both latency and communication overhead.

\section{CONCLUSION}
\label{sec:conclusion}
In this paper, we propose FastQuery, a private embedding table query framework, which leverages the one-hot nature of queries and robustness to
low bit-width quantization noise, to improve communication efficiency.
FastQuery features a communication-aware embedding table quantization algorithm and a one-hot-aware dense packing algorithm. 
Compared to prior-art HE-based frameworks, e.g., Cheetah, Iron, and Bumblebee, FastQuery achieves more than $4.3\times$, $2.7\times$, $1.3\times$ latency reduction,
respectively, and more than $75.7\times$, $60.2\times$, $20.2\times$ communication reduction, respectively, on both LLAMA-7B and LLAMA-30B.

\section*{Acknowledge}
This work was supported in part by the NSFC (62125401) and the 111 Project (B18001). We also thank Prof. Yier Jin and Dr. Jiafeng Hua from Huawei Co. Ltd. for the useful discussion and support.

{
\small
\bibliographystyle{plain}
\bibliography{ref}
}

\end{document}